\documentclass[conference]{IEEEtran}
\IEEEoverridecommandlockouts
\usepackage{cite}
\usepackage{amsmath,amssymb,amsfonts}
\usepackage{amsmath}
\usepackage{stfloats}   

\usepackage{algorithm}      
\usepackage{algpseudocode}  

\usepackage{graphicx}
\usepackage{textcomp}
\usepackage{xcolor}
\def\BibTeX{{\rm B\kern-.05em{\sc i\kern-.025em b}\kern-.08em
    T\kern-.1667em\lower.7ex\hbox{E}\kern-.125emX}}

\graphicspath{{figs/}}

\begin{document}
\title{DRL-Based Resource Allocation for Energy-Efficient IRS-Assisted UAV Spectrum Sharing Systems}

\author{
\IEEEauthorblockN{
Yiheng Wang\IEEEauthorrefmark{1}\IEEEauthorrefmark{2}\\
\IEEEauthorblockA{\IEEEauthorrefmark{1}School of Information Science and Technology, Beijing University of Technology, Beijing, 100124, China}
\IEEEauthorblockA{\IEEEauthorrefmark{2}Faculty of Electrical and Electronic Engineering, University of Nottingham, Nottingham, NG7 2RD, United Kingdom}
\IEEEauthorblockA{Emails: 22020025@emails.bjut.edu.cn; eeyyw12@nottingham.ac.uk}
}}

\maketitle

\begin{abstract}
Intelligent reflecting surface (IRS)-assisted unmanned aerial vehicle (UAV) systems provide a new paradigm for reconfigurable and flexible wireless communications. To enable more energy-efficient and spectrum-efficient IRS-assisted UAV wireless communications, this paper introduces a novel IRS-assisted UAV-enabled spectrum sharing system with orthogonal frequency-division multiplexing (OFDM). The goal is to maximize the energy efficiency (EE) of the secondary network by jointly optimizing the beamforming, subcarrier allocation, IRS phase shifts, and the UAV trajectory subject to practical transmit-power and passive-reflection constraints as well as UAV physical limitations. A physically grounded propulsion-energy model is adopted, with its tight upper bound used to form a tractable EE lower bound for the spectrum sharing system. To handle highly non-convex, time-coupled optimization problems with a mixed continuous and discrete policy space, we develop a deep reinforcement learning (DRL) approach based on the actor–critic framework. Extended experiments show the significant EE improvement of the proposed DRL-based approach compared to several benchmark schemes, thus demonstrating the effectiveness and robustness of the proposed approach with mobility.
\end{abstract}


\section{Introduction}
Unmanned aerial vehicles (UAVs) are emerging as a key enabler of next-generation wireless systems due to their high mobility and favorable line-of-sight (LoS) connectivity \cite{9866052}. By serving as aerial relays or access points, UAVs can provide rapid coverage extension and flexible deployment in scenarios where terrestrial infrastructure is limited \cite{bacsturk2022energy}. However, UAVs-enabled communication systems are fundamentally constrained by limited onboard batteries \cite{wang2024distributed}. Since propulsion energy consumption far exceeds communication-related power, the energy efficiency (EE) becomes the dominant performance metric \cite{10660499}. Designing communication systems that explicitly account for UAV propulsion models is therefore crucial for sustainable operations \cite{ning20256g}.

In addition to energy concerns, UAV-based communication networks also face the challenge of spectrum efficiency (SE) \cite{xie2025resource}. Supporting multiple users over broadband orthogonal frequency-division multiplexing (OFDM) systems requires careful resource allocation to handle frequency-selective fading, inter-subcarrier interference, and dynamic user distributions \cite{10557531}. Moreover, UAV links often operate in spectrum-sharing systems with terrestrial networks, which introduces additional interference management challenges and necessitates efficient spectrum reuse strategies \cite{qin2023deep}. Without coordinated beamforming, scheduling, and sharing policies, the available spectral resources can not be fully exploited, limiting the overall throughput gains \cite{kieu2024uav}.

To address these dual challenges of energy and spectrum efficiency, intelligent reflecting surfaces (IRSs) have recently attracted significant interest \cite{wang2023intelligent}. By leveraging large arrays of passive reflection elements with programmable phase shifts, IRSs can reshape the wireless propagation environment at a low power cost \cite{10559487}. Mounting an IRS on a UAV enables both mobility and reconfigurability, creating controllable cascaded links between base stations and ground users. This synergy offers new opportunities to jointly enhance EE and SE, but also introduces highly coupled design variables spanning UAV trajectories, IRS phase configurations, and multi-carrier beamforming.

Conventional optimization techniques, such as alternating optimization (AO) and successive convex approximation (SCA), cannot handle the resulting non-convex and time-dependent formulations in a real-time manner \cite{10806777}. To overcome these limitations, deep reinforcement learning (DRL) has emerged as a promising approach \cite{10319408}. By interacting with the environment and learning policies through trial-and-error, DRL can adapt to highly dynamic wireless conditions and long-term performance objectives \cite{11121573}. Unlike traditional model-based methods, DRL is capable of capturing the intricate trade-offs between energy efficiency, spectrum utilization, and spectrum sharing, while maintaining robustness under user mobility and channel uncertainties \cite{xu2024learning}.

In this paper, we investigate an IRS-assisted UAV-assisted spectrum sharing system with OFDM, and maximize the EE of the secondary network by jointly optimizing SBS beamforming, IRS passive reflection, and UAV trajectory while maintaining the transmission quality of the primary network. We discretize the mission into time slots, derive closed-form signal-to-interference-plus-noise ratio (SINR) and rate expressions per subcarrier, and adopt a physically grounded propulsion energy model to form an EE objective for stable training. Building on these models, we design a constrained actor–critic DRL framework that performs the mixed discrete-continuous actions while respecting per-slot power, passive-reflection, and kinematic constraints. Extended simulation results demonstrate the significant EE improvement of the proposed DRL-based method compared to the baselines, and show robustness to mobility dynamics.

\section{System Model and Problem Formulation}\label{sec:sys_prob}
\begin{figure}[t!]
    \centering
    \includegraphics[width=1\linewidth]{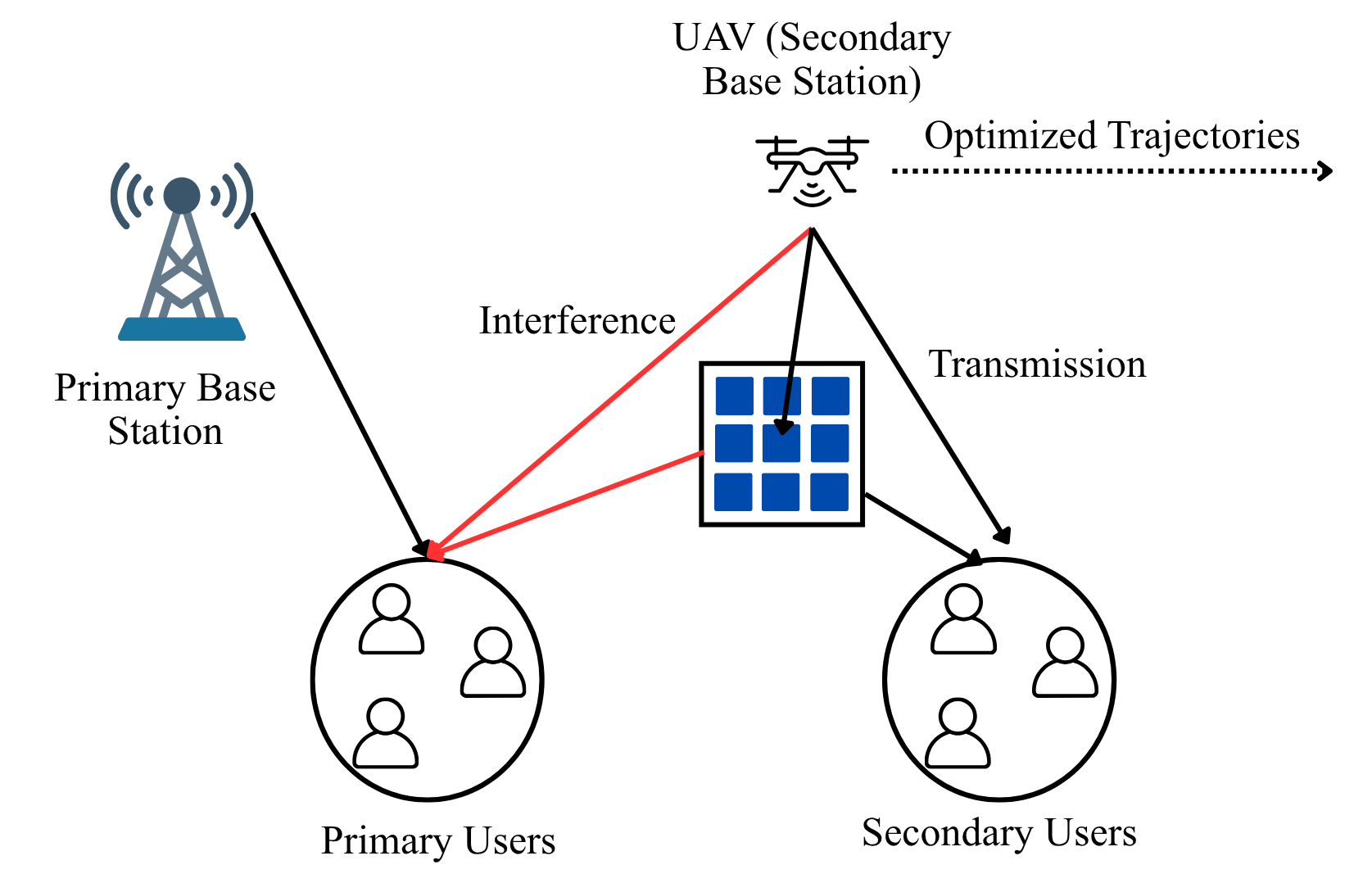}
    \caption{The considered system model of UAV-IRS spectrum sharing systems.}
    \vspace{-0.5cm}
    \label{fig:1}
\end{figure}

\subsection{System Model}\label{sec:system_model}
We consider a spectrum-sharing downlink OFDM communication system assisted by a UAV-mounted intelligent reflecting surface (IRS), as illustrated in Fig.~1. 
The system consists of a multi-antenna primary base station (PBS), a UAV-mounted secondary base station (SBS) equipped with an IRS, 
and two groups of single-antenna ground users: the set of primary users (PUs) denoted by $\mathcal{D}=\{1,\ldots,D_p\}$ 
and the set of secondary users (SUs) denoted by $\mathcal{K}=\{1,\ldots,K\}$.  
The PBS has $N_t$ antennas, while the IRS comprises $N_I$ passive reflecting elements mounted on the UAV, whose flight trajectory can be controlled.  
The total system bandwidth is $B$~Hz and is divided into $D$ OFDM subcarriers, each with bandwidth $B/D$.  
Both PBS and UAV-SBS share the same spectrum resources, which may lead to co-channel interference when PUs and SUs are scheduled on the same subcarrier.  

The IRS applies element-wise reflections to reconfigure the propagation environment.  
Let
\begin{equation}
    \boldsymbol{\Phi}=\mathrm{diag}(\phi_1,\ldots,\phi_{N_I}) \in \mathbb{C}^{N_I\times N_I},
\end{equation}
denote the IRS phase-shift matrix, where $\phi_n=\beta_n e^{j\theta_n}$ is the reflection coefficient of element $n$, 
$\beta_n\in[0,1]$ (typically $\beta_n=1$), and $\theta_n\in[0,2\pi)$.  
Let $\mathbf{H}_{p,R}\in\mathbb{C}^{N_I\times N_t}$ denote the channel from the PBS to the IRS, 
$\mathbf{h}_{r,k}\in\mathbb{C}^{N_I\times 1}$ the channel from the IRS to SU $k$, 
$\mathbf{h}_{p,d}\in\mathbb{C}^{N_t\times 1}$ the direct channel from the PBS to PU $d$, 
and $\mathbf{g}_{s,k}\in\mathbb{C}^{N_t\times 1}$ the channel from the UAV-SBS (via IRS) to SU $k$.  
For subcarrier $d\in\{1,\ldots,D\}$, let $\mathbf{w}_{d}\in\mathbb{C}^{N_t\times 1}$ be the PSBS beamforming vector for PU $d$, 
and $\mathbf{v}_{k}\in\mathbb{C}^{N_t\times 1}$ the SSBS beamforming vector for SU $k$.  
Let $s_{d}\sim\mathcal{CN}(0,1)$ and $s_{k}\sim\mathcal{CN}(0,1)$ denote the transmit symbols of PU $d$ and SU $k$, respectively.  

The received signal at PU $d$ on subcarrier $d$ can be represented by
\begin{equation}
    y_{p,d}
    =
    \mathbf{h}_{p,d}^{H}\mathbf{w}_{d}\,s_{d}
    + \underbrace{\sum_{k\in\mathcal{K}} \mathbf{h}_{r,d}^{H}\boldsymbol{\Phi}\mathbf{v}_{k}\,s_{k}}_{\text{interference from SUs}}
    + n_{d},
\end{equation}
where $n_{d}\sim\mathcal{CN}(0,\sigma_{p,d}^{2})$ is AWGN. Moreover, the received signal at SU $k$ on subcarrier $d$ is represented by 
\begin{equation}
    y_{k,d}
    =
    \big(\mathbf{g}_{s,k}^{H} + \mathbf{h}_{r,k}^{H}\boldsymbol{\Phi}\mathbf{H}_{p,R}\big)\mathbf{v}_{k}\,s_{k}
    + \underbrace{\sum_{d\in\mathcal{D}} \mathbf{h}_{p,k}^{H}\mathbf{w}_{d}\,s_{d}}_{\text{interference from PUs}}
     + n_{k},
\end{equation}
where 
$n_{k}\sim\mathcal{CN}(0,\sigma_{k,d}^{2})$ is AWGN.  
The per-subcarrier noise variance is
\begin{equation}
    \sigma_{u,d}^{2}=N_{0}\,\frac{B}{D}\,F_{u},
\end{equation}
where $N_0$ is the noise power spectral density (W/Hz) and $F_u$ the noise figure of user $u$ in linear scale.  

The power constraint of the UAV-SBS is expressed as
\begin{equation}
\sum_{k\in\mathcal{K}}\sum_{d=1}^{D}\|\mathbf{v}_{k}\|^{2} \le P_{s,\max},
|\phi_n| \le 1, \quad \theta_n\in[0,2\pi),
\end{equation}
where $(\cdot)^H$ denotes the Hermitian transpose, $\|\cdot\|$ denotes the Euclidean norm, $|\cdot|$ denotes the complex modulus, and $P_{s,\max}$ is the maximum transmit power of the UAV-SBS.

\subsection{Problem Formulation}\label{sec:problem}

We discretize the UAV mission into $N$ time intervals of duration $\delta_t$.
In slot $n$, the UAV state is position $\mathbf{q}[n]$, velocity $\mathbf{v}[n]$ and acceleration $\mathbf{a}[n]$,
where $\mathbf{q}[n]\in\mathbb{R}^3$, $\mathbf{v}[n]\in\mathbb{R}^3$, and $\mathbf{a}[n]\in\mathbb{R}^3$ denote the UAV position, velocity, and acceleration vectors, respectively.
The channels may vary with $n$; when necessary, we write $\mathbf{H}_{b,R}[n]$, $\mathbf{h}_{r,k}[n]$, $\mathbf{h}_{d,k}[n]$, $\boldsymbol{\Phi}[n]$, $\mathbf{w}_{k,d}[n]$.

Define the effective row channel:
\begin{equation}
\label{eq:geff}
    \mathbf{g}_{k}^{H}[n]=\mathbf{h}_{d,k}^{H}[n]+\mathbf{h}_{r,k}^{H}[n]\boldsymbol{\Phi}[n]\mathbf{H}_{b,R}[n]\in\mathbb{C}^{1\times N_t}.
\end{equation}
Here, $\mathbf{g}_{k}^{H}[n]$ aggregates the direct and reflected channels from the BS to user $k$.
Then, the SINR and the rate in the unit of bit/s of user $k$ on subcarrier $d$ at time slot $n$ are, respectively, represented by
\begin{equation}\label{eq:sinr_rate}
\gamma_{k,d}[n] =
\frac{\big|\mathbf{g}_k^{H}[n]\mathbf{w}_{k,d}[n]\big|^{2}}
{\displaystyle\sum_{i\in\mathcal{K},\,i\neq k}\big|\mathbf{g}_k^{H}[n]\mathbf{w}_{i,d}[n]\big|^{2}+\sigma_{k,d}^{2}},
\end{equation}
and
\begin{equation}
R_{k,d}[n] = \tfrac{B}{D}\log_{2}\!\big(1+\gamma_{k,d}[n]\big),
\end{equation}
where $\gamma_{k,d}[n]$ is the instantaneous SINR and $R_{k,d}[n]$ the achievable rate of user $k$.
The total bits transmitted over the horizon $N$ of the IRS-UAV secondary network is expressed by
\begin{equation}\label{eq:Rtotal}
R^{\text{total}}=\sum_{n=1}^{N}\sum_{d=1}^{D}\sum_{k=1}^{K} R_{k,d}[n]\delta_t ,
\end{equation}
where $R^{\text{total}}$ denotes the accumulated transmitted bits during the mission.

\begin{figure*}[!t]
\begin{equation}\label{eq:Etot}
E^{\text{total}}
= \delta_t \sum_{n=1}^{N}\!\left[
c_{1}\,\|\mathbf{v}[n]\|^{3}
+\frac{c_{2}}{\|\mathbf{v}[n]\|}\!
\left(1+\frac{\|\mathbf{a}[n]\|^{2}-\dfrac{(\mathbf{a}^{T}[n]\mathbf{v}[n])^{2}}{\|\mathbf{v}[n]\|^{2}}}{g^{2}}\right)
\right]
+\tfrac{1}{2}m\!\left(\|\mathbf{v}[N]\|^{2}-\|\mathbf{v}[0]\|^{2}\right),
\end{equation}
\hrulefill
\vspace{-0.8em}
\end{figure*}

The total energy consumption can be represented by the propulsion-energy model \eqref{eq:Etot}, where $c_1$ and $c_2$ are UAV aerodynamic parameters and $g$ is the gravitational acceleration constant, and $R^{\text{total}}$ denotes the accumulated transmitted bits during the mission. Then, the total energy consumption can be upper-bounded for tractability by \cite{9866052}
\begin{equation}\label{eq:Eub}
E^{\text{total}} \le E^{\text{total}}_{\mathrm{ub}}
= \delta_t \sum_{n=1}^{N}\!\left[
c_{1}\|\mathbf{v}[n]\|^{3}
+\frac{c_{2}}{\|\mathbf{v}[n]\|}\!\left(1+\frac{\|\mathbf{a}[n]\|^{2}}{g^{2}}\right)
\right],
\end{equation}
where $m$ denotes the UAV mass including IRS payload.
The lower-bounded energy efficiency (EE) is defined as
\begin{equation}
    \mathrm{EE}_{\mathrm{lb}}=\frac{R^{\text{total}}}{E^{\text{total}}_{\mathrm{ub}}}.
\end{equation}

In the considered spectrum-sharing scenario, the system objective is to maximize the overall energy efficiency (EE), defined as the ratio between the accumulated transmitted bits of both the primary and secondary networks and the total energy consumption over the UAV trajectories. The optimization problem is formulated by
\begin{equation}
\begin{aligned}
&\max_{\{\mathbf{w}_{k,d}[n]\},\,\{\boldsymbol{\Phi}[n]\},\,\{\mathbf{q}[n],\mathbf{v}[n],\mathbf{a}[n]\}}
\quad \mathrm{EE}_{\mathrm{lb}}\\
\text{s.t.}& \text{(C1): } \sum_{d=1}^{D}\sum_{k=1}^{K}\|\mathbf{w}_{k,d}[n]\|^{2} \le P_{\max},\ \forall n, \\
& \text{(C2): }\boldsymbol{\Phi}[n]=\mathrm{diag}\big(\phi_{1}[n],\ldots,\phi_{N_I}[n]\big), \ |\phi_{\ell}[n]|\le 1,\ \forall \ell,n, \\
& \text{(C3): }\mathbf{q}[n{+}1]=\mathbf{q}[n]+\mathbf{v}[n]\delta_t, 
                         \mathbf{v}[n{+}1]=\mathbf{v}[n]+\mathbf{a}[n]\delta_t, \\
& \text{(C4): }\|\mathbf{v}[n]\|\le v_{\max}, \|\mathbf{a}[n]\|\le a_{\max},\ \forall n, \\
& \text{(C5): }\mathbf{q}[0]=\mathbf{q}_0,\quad \mathbf{q}[N]=\mathbf{q}_F, \|\mathbf{v}[n]\|\ge v_{\min}>0, \\
& \text{(C6): }\sum_{k=1}^{K}\big|\mathbf{h}_{r,d}^{H}\boldsymbol{\Phi}\mathbf{v}_{k}\mathbf{w}_{k,d}[n]\big|^{2} \le \Gamma_{d}, \forall d,n,
\end{aligned}
\end{equation}
where (C1) represents the BS transmit power constraint,
(C2) represents the IRS unit module constraint,
(C3) represents the kinematic update of the UAV, (C4) enforces the limits of velocity and acceleration, 
(C5) constrains the initial and final positions of the UAV together with a minimum speed requirement,
and (C6) represents the maximum interface from the secondary network to the primary network.

\section{DRL-Based Energy-Efficient Transmission Framework}\label{sec:drl}

Given the non-convexity and time coupling of the EE maximization problem formulated in (13), we develop a hybrid deep reinforcement learning (DRL) framework that factorizes control into two coupled components: (i) discrete OFDM subcarrier-user scheduling, and (ii) continuous radio-and-motion control. These are handled by coordinated heads that share a common state encoder, i.e., a dueling double deep Q-network (D3QN) head for discrete scheduling, and a Soft actor-critic (SAC) head for continuous variables. 

\subsection{MDP Formulation}\label{sec:mdp}
\textbf{State Space.} At slot $t$, the observable state aggregates both communication channels and UAV kinematics:
\begin{equation}
\begin{aligned}
\label{eq:state}
\small
\mathbf{s}[t]=&\Big\{
\mathbf{H}_{b,R}[t],\{\mathbf{h}_{r,k}[t]\}_{k\in\mathcal{K}},\{\mathbf{h}_{d,k}[t]\}_{k\in\mathcal{K}},
\{\sum_{k\in\mathcal{K}} \mathbf{h}_{r,d}^{H}\boldsymbol{\Phi}\mathbf{v}_{k}\}_{d}, \\
&\boldsymbol{\Phi}[t-1],\mathbf{q}[t],\mathbf{v}[t],\mathbf{a}[t],
\overline{\mathrm{EE}}[t-1]
\Big\},
\end{aligned}
\end{equation}
where $\{\sum_{k\in\mathcal{K}} \mathbf{h}_{r,d}^{H}\boldsymbol{\Phi}\mathbf{v}_{k}\}_{d}$ denotes the effective interference channels from the UAV-SBS to the PUs, which captures limits from constraint (C6), and $\overline{\mathrm{EE}}[t-1]$ is an exponential moving average of per-slot EE to stabilize training.

\textbf{Action Space.}
The action space is factorized as
\begin{equation}
\label{eq:action_fact}
\mathcal{A}=\mathcal{A}_{\text{disc}}\times \mathcal{A}_{\text{cont}},
\end{equation}
where $\mathcal{A}_{\text{disc}}$ handles subcarrier scheduling and $\mathcal{A}_{\text{cont}}$ handles continuous beamforming, IRS phases, and UAV control.
The discrete action \,$\mathcal{A}_{\text{disc}}$ from the D3QN head is 
\begin{equation}
\label{eq:disc}
x_{k,d}[t]\in\{0,1\},\quad \sum_{k=0}^{K}x_{k,d}[t]=1,\ \forall d,
\end{equation}
where $x_{k,d}[t]=0$ denotes an idle subcarrier assignment. Collectively, $\mathbf{X}[t]=[x_{k,d}[t]]\in\{0,1\}^{(K+1)\times D}$.
The continuous action \,$\mathcal{A}_{\text{cont}}$ from the SAC head is
\begin{equation}
\label{eq:cont}
\mathbf{a}_{\text{cont}}[t]=\Big\{\{\mathbf{w}_{k,d}[t]\}_{k,d},\ \boldsymbol{\Phi}[t],\ \mathbf{a}[t]\Big\},
\end{equation}
where $\mathbf{w}_{k,d}[t]\in\mathbb{C}^{N_t}$ are UAV-SBS beamformers, $\boldsymbol{\Phi}[t]=\mathrm{diag}(\phi_1[t],\dots,\phi_{N_I}[t])$ denotes IRS reflection coefficients, and $\mathbf{a}[t]$ is the UAV acceleration vector. Velocity and position evolve per kinematics (C3).

Channels evolve according to UAV geometry and LoS propagation. UAV motion evolves as $\mathbf{q}[t+1]=\mathbf{q}[t]+\mathbf{v}[t]\delta_t$, $\mathbf{v}[t+1]=\mathbf{v}[t]+\mathbf{a}[t]\delta_t$. The scheduled sum rate is
\begin{equation}
\label{eq:sched_sumrate}
S[t]=\sum_{d=1}^{D}\sum_{k=1}^{K} x_{k,d}[t] R_{k,d}[t],
\end{equation}
where $R_{k,d}[t]$ follows \eqref{eq:sinr_rate}.

\subsection{Reward Design with Explicit EE Constraint Penalties}\label{sec:reward}
The per-slot reward adopts an EE-proxy defined as the scheduled sum rate divided by the propulsion-energy integrand of \eqref{eq:Eub} and penalized by constraints (C1)-(C6).

First of all, the propulsion-energy integrand is
\begin{equation}
\label{eq:eprop}
e_{\text{prop}}[t]=
c_1\|\mathbf{v}[t]\|^{3}
+\frac{c_2}{\|\mathbf{v}[t]\|}\!\left(1+\frac{\|\mathbf{a}[t]\|^{2}}{g^{2}}\right).
\end{equation}
We further design the constraint penalties corresponding to (C1)-(C6), each of which quantifies the normalized violation of a specific feasibility condition. These penalty functions are incorporated into the reward so that the agent learns to respect system and physical constraints. Specifically, for the BS transmit power constraint, the following penalty measures the relative violation of the maximum transmit power $P_{\max}$:
\begin{equation}
\label{eq:Gpow}
G_{\text{pow}}[t]=\frac{\Big[\sum_{k,d}\|\mathbf{w}_{k,d}[t]\|^{2}-P_{\max}\Big]_+}{P_{\max}}.
\end{equation} 
To enforce $|\phi_\ell|=1$ for each IRS element, the penalty is defined as the average deviation of reflection coefficients by
\begin{equation}
\label{eq:Girs}
G_{\text{irs}}[t]=\frac{1}{N_I}\sum_{\ell=1}^{N_I}\big[\,|\phi_\ell[t]|-1\,\big]_+.
\end{equation}
Moreover, we design the penalty accounts for exceeding the maximum speed $v_{\max}$ or falling below the minimum admissible speed $v_{\min}$:
\begin{equation}
\label{eq:Gspd}
G_{\text{spd}}[t]=\frac{[\|\mathbf{v}[t]\|-v_{\max}]_+}{v_{\max}}
+\frac{[v_{\min}-\|\mathbf{v}[t]\|]_+}{v_{\min}}.
\end{equation} 
Violations of the UAV acceleration limit $a_{\max}$ are penalized as
\begin{equation}
\label{eq:Gacc}
G_{\text{acc}}[t]=\frac{[\|\mathbf{a}[t]\|-a_{\max}]_+}{a_{\max}}.
\end{equation}
To ensure spectrum sharing feasibility, the interference leakage from the UAV-SBS towards each PU on subcarrier $d$ is constrained below $\Gamma_d$. The violation is penalized as
\begin{equation}
\label{eq:Gintf}
G_{\text{intf}}[t]=\frac{\Big[\sum_{k=1}^{K}\big|\mathbf{h}_{r,d}^{H}\boldsymbol{\Phi}\mathbf{v}_{k}\mathbf{w}_{k,d}[n]\big|^{2}-\Gamma_d\Big]_+}{\Gamma_d}.
\end{equation}
At the end of the mission horizon, the UAV is required to reach the final waypoint $\mathbf{q}_F$. A normalized distance to the target is imposed as a terminal penalty by
\begin{equation}
\label{eq:Gterm}
G_{\text{term}}=\frac{\|\mathbf{q}[N]-\mathbf{q}_F\|}{\|\mathbf{q}_F-\mathbf{q}_0\|+\varepsilon}.
\end{equation}
Here, $[u]_+=\max(u,0)$ ensures nonnegativity, and $\varepsilon>0$ prevents division by zero.

Hence, the per-slot reward can be represented by
\begin{equation}
\begin{aligned}
\label{eq:reward}
r_t&=\frac{S[t]}{e_{\text{prop}}[t]+\varepsilon}
-\beta_1 G_{\text{pow}}[t]-\beta_2 G_{\text{irs}}[t] \\
&-\beta_3 G_{\text{spd}}[t]-\beta_4 G_{\text{acc}}[t]
-\beta_5 G_{\text{intf}}[t],
\end{aligned}
\end{equation}
with a terminal shaping $-\beta_6 G_{\text{term}}$, and the coefficients $\{\beta_i\}>0$ weight the penalties.

\subsection{Policy Parameterization and Analytic Projections}\label{sec:param_proj}

To handle the mixed discrete–continuous decision variables, we adopt a dual-head DRL architecture where the discrete scheduling actions are modeled by a Dueling Double Deep Q-Network (D3QN), and the continuous variables are parameterized by a Soft Actor–Critic (SAC) policy. A shared state encoder extracts channel and mobility features, which are then fed into the two specialized heads.

For the discrete head, let $Q_\theta(\mathbf{s},a_{\text{disc}})$ denote the action–value function of state $\mathbf{s}$ under discrete action $a_{\text{disc}}$, parameterized by $\theta$. The D3QN decomposes $Q_\theta$ into a state-value stream $V_\theta(\mathbf{s})$ and an advantage stream $A_\theta(\mathbf{s},a_{\text{disc}})$ as
\begin{equation}
Q_\theta(\mathbf{s},a_{\text{disc}})=V_\theta(\mathbf{s})+\Big(A_\theta(\mathbf{s},a_{\text{disc}})-\tfrac{1}{|\mathcal{A}_{\text{disc}}|}\sum_{a'}A_\theta(\mathbf{s},a')\Big),
\end{equation}
where the subtraction enforces identifiability. The training target is given by the double Q-learning update
\begin{equation}
y_{\text{D}}=r+\gamma\,Q_{\theta^-}\!\big(\mathbf{s}',\arg\max_{a'}Q_\theta(\mathbf{s}',a')\big),
\end{equation}
where $\theta^-$ are target-network parameters updated by Polyak averaging. The loss function is
\begin{equation}
\mathcal{L}_{\text{D3QN}}=\mathbb{E}\Big[\big(y_{\text{D}}-Q_\theta(\mathbf{s},a_{\text{disc}})\big)^2\Big].
\end{equation}
For each subcarrier $d$, logits over $\{0,\dots,K\}$ are generated, where $k=0$ denotes idle allocation. Invalid assignments are excluded by action masking.

For the continuous head, we employ SAC. Let $\pi_\lambda(\mathbf{a}_{\text{cont}}|\mathbf{s})$ denote the stochastic Gaussian policy parameterized by $\lambda$, where the actor outputs mean $\mu_\lambda(\mathbf{s})$ and log-variance $\sigma^2_\lambda(\mathbf{s})$. The sampled action is
\begin{equation}
\mathbf{a}_{\text{cont}}=\tanh(\mu_\lambda(\mathbf{s})+\sigma_\lambda(\mathbf{s})\odot \xi),\quad \xi\sim\mathcal{N}(0,I),
\end{equation}
which is then mapped into beamformers $\{\mathbf{w}_{k,d}\}$, IRS coefficients $\boldsymbol{\Phi}$, and UAV acceleration $\mathbf{a}$. The critics $Q_{\omega_1},Q_{\omega_2}$ are trained with the soft Bellman target
\begin{equation}
y_{\text{S}}=r+\gamma\Big(\min_{j=1,2}Q_{\omega_j^-}(\mathbf{s}',\mathbf{a}')-\alpha \log\pi_\lambda(\mathbf{a}'|\mathbf{s}')\Big), \mathbf{a}'\sim\pi_\lambda(\cdot|\mathbf{s}'),
\end{equation}
and the critic loss is
\begin{equation}
\mathcal{L}_{\text{critic}}=\sum_{j=1}^{2}\mathbb{E}\Big[\big(y_{\text{S}}-Q_{\omega_j}(\mathbf{s},\mathbf{a}_{\text{cont}})\big)^2\Big].
\end{equation}
The actor is updated by minimizing
\begin{equation}
\mathcal{L}_{\text{actor}}=\mathbb{E}\Big[\alpha \log\pi_\lambda(\mathbf{a}_{\text{cont}}|\mathbf{s})-\min_{j}Q_{\omega_j}(\mathbf{s},\mathbf{a}_{\text{cont}})\Big],
\end{equation}
while the temperature parameter $\alpha$ is adapted via
\begin{equation}
\mathcal{L}_{\alpha}=\mathbb{E}\big[-\alpha(\log\pi_\lambda(\mathbf{a}_{\text{cont}}|\mathbf{s})+\bar{H})\big],
\end{equation}
where $\bar{H}$ is the target entropy.

To ensure feasibility, analytic projections are applied after sampling the continuous actions. Beamformers are rescaled to meet the BS power constraint (C1), IRS coefficients are projected onto the unit circle to satisfy (C2), and UAV velocities and accelerations are projected to enforce (C3)-(C4). The terminal waypoint constraint (C5) and the interference-protection constraint (C6) cannot be analytically projected and are therefore incorporated as reward penalties by (26). 

\subsection{Learning Objective}\label{sec:loss}
The overall objective is to maximize the expected discounted return, which is represented by
\begin{equation}
J=\mathbb{E}\Big[\sum_{t=0}^{\infty}\gamma^{t}r_t\Big],\quad \gamma\in(0,1],
\end{equation}
where $r_t$ is the per-slot reward in \eqref{eq:reward}. The hybrid architecture thus jointly optimizes a discrete scheduler and a continuous controller by combining D3QN-based Q-learning for $\mathcal{A}_{\text{disc}}$ and SAC-based policy gradient optimization for $\mathcal{A}_{\text{cont}}$. This formulation allows the UAV-assisted IRS system to learn energy-efficient strategies in a spectrum-sharing environment with stringent power, mobility, and interference constraints.

\subsection{Overall Algorithm}\label{sec:algo}
The overall hybrid SAC--D3QN with analytic projections operates by encoding the state, factorizing the action into discrete scheduling and continuous controls, applying projections for (C1)-(C5), and incorporating interference protection (C6) through reward penalties. This framework ensures robust energy-efficient transmission under realistic UAV mobility and spectrum-sharing conditions. The proposed DRL-based approach is concluded in Algorithm 1.

\begin{algorithm}[t!]
\caption{Hybrid D3QN--SAC with Projections for EE-Optimal Spectrum Sharing}
\label{alg:hybrid_sac_d3qn}
\begin{algorithmic}[1]
\small
\State \textbf{Inputs:} discount $\gamma$, learning rates $\eta_Q,\eta_V,\eta_\pi,\eta_\alpha$, target update rate $\tau$, entropy target $\bar H$, initial penalties $\{\beta_i\}_{i=1}^{6}$, episode length $N$, batch size $B$, gradient steps per slot $G$
\State \textbf{Initialize:} shared encoder $\mathrm{Enc}_{\psi}$; \hspace{0.2em}D3QN online head $Q_{\theta}$ and target $Q_{\theta^-}\!\leftarrow\!\theta$; \hspace{0.2em}SAC critics $Q_{\omega_1},Q_{\omega_2}$ and targets $Q_{\omega_1^-}\!\leftarrow\!\omega_1$, $Q_{\omega_2^-}\!\leftarrow\!\omega_2$; \hspace{0.2em}actor $\pi_{\lambda}$; temperature $\alpha>0$; replay buffer $\mathcal{D}$
\For{episode $e=1,2,\dots$}
  \State Reset environment and obtain initial state $\mathbf{s}_0$
  \For{slot $t=0,1,\dots,N{-}1$}
    \State Encode features: $\mathbf{z}_t \!\leftarrow\! \mathrm{Enc}_{\psi}(\mathbf{s}_t)$
    \State $\tilde{\mathbf{u}}_t \sim \mathcal{N}\!\big(\mu_\lambda(\mathbf{z}_t),\ \Sigma_\lambda(\mathbf{z}_t)\big),\quad
           \mathbf{a}^{\text{raw}}_{\text{cont},t} = \tanh(\tilde{\mathbf{u}}_t)$
    \State Apply projections for (C1)-(C4)
    \State Execute $(\mathbf{X}_t,\{\mathbf{w}_{k,d}\}_t,\boldsymbol{\Phi}_t,\mathbf{a}_t)$ for the next state $\mathbf{s}_{t+1}$
    \State Compute scheduled sum-rate $S[t]$ via \eqref{eq:sched_sumrate}
    \State Compute $e_{\text{prop}}[t]$ via \eqref{eq:eprop} and penalties
    \State Compute reward $r_t=\frac{S[t]}{e_{\text{prop}}[t]+\varepsilon} - \sum_{i=1}^{5}\beta_i\,G_i[t]$
    \State Store transition $(\mathbf{s}_t,\ \mathbf{X}_t,\ \mathbf{a}^{\text{raw}}_{\text{cont},t},\ r_t,\ \mathbf{s}_{t+1})$ into $\mathcal{D}$
    \For{$g=1$ to $G$} \Comment{Off-policy updates}
      \State Sample minibatch $\{(\mathbf{s},\mathbf{X},\mathbf{a}^{\text{raw}}_{\text{cont}},r,\mathbf{s}')\}_{b=1}^{B}\!\sim\!\mathcal{D}$
      \State Encode $\mathbf{z}\!=\!\mathrm{Enc}_{\psi}(\mathbf{s})$, $\mathbf{z}'\!=\!\mathrm{Enc}_{\psi}(\mathbf{s}')$
      \State Compute $y_{\text{D}} = r + \gamma\, Q_{\theta^-}\big(\mathbf{z}',\arg\max_{a'} Q_{\theta}(\mathbf{z}',a')\big)$
      \State Update $\theta \leftarrow \theta - \eta_Q \nabla_{\theta}\, \frac{1}{B}\!\sum_b \big(y_{\text{D}} - Q_{\theta}(\mathbf{z},a_{\text{disc}})\big)^2$
      \State Sample $\mathbf{a}'\!\sim\!\pi_{\lambda}(\cdot|\mathbf{z}')$
      \State Compute $y_{\text{S}} = r + \gamma\big(\min_{j=1,2}Q_{\omega_j^-}(\mathbf{z}',\mathbf{a}') - \alpha \log \pi_{\lambda}(\mathbf{a}'|\mathbf{z}')\big)$
      \State Critic update: $\omega_j \leftarrow \omega_j - \eta_V \nabla_{\omega_j}\, \frac{1}{B}\!\sum_b \big(y_{\text{S}} - Q_{\omega_j}(\mathbf{z},\mathbf{a}_{\text{cont}})\big)^2,\ j\in\{1,2\}$
      \State Actor update: $\lambda \leftarrow \lambda - \eta_\pi \nabla_{\lambda}\, \frac{1}{B}\!\sum_b \Big(\alpha \log \pi_{\lambda}(\mathbf{a}_{\text{cont}}|\mathbf{z}) - \min_{j}Q_{\omega_j}(\mathbf{z},\mathbf{a}_{\text{cont}})\Big)$
      \State Temperature update: $\alpha \leftarrow \alpha - \eta_\alpha \nabla_{\alpha}\, \frac{1}{B}\!\sum_b \big(-\alpha(\log\pi_{\lambda}(\mathbf{a}_{\text{cont}}|\mathbf{z})+\bar H)\big)$
      \State Target soft updates:
             $\theta^- \leftarrow \tau \theta + (1-\tau)\theta^-$ and 
             $\omega_j^- \leftarrow \tau \omega_j + (1-\tau)\omega_j^-$,\ $j\!\in\!\{1,2\}$
    \EndFor
    \State $\mathbf{s}_t \leftarrow \mathbf{s}_{t+1}$
  \EndFor
\EndFor
\end{algorithmic}
\end{algorithm}

\section{Simulation and Results}\label{sec:sim}
In this section, we evaluate the performance of the proposed DRL-based energy-efficient resource allocation for the IRS-assisted UAV spectrum sharing network.

\subsection{Simulation Setup}
The carrier frequency is set to $f_c=2.5$~GHz, with total bandwidth $B=10$~MHz equally divided into $D=64$ OFDM subcarriers. The noise power spectral density is $N_0=-174$~dBm/Hz. The BS is equipped with $N_t=4$ transmit antennas, and the IRS mounted on the UAV contains $N_I=64$ reflecting elements. The UAV flies at a fixed altitude of $H=100$~m with mission duration $T=40$~s, discretized into $N=40$ time slots. The UAV maximum velocity and acceleration are $v_{\max}=20$~m/s and $a_{\max}=5$~m/s$^2$, respectively, with a minimum velocity $v_{\min}=3$~m/s. The maximum transmit power of the BS is $P_{\max}=30$~dBm. The interference threshold for protecting the primary network is $\Gamma_d=-90$~dBm for each subcarrier $d$. We compare the proposed DRL method with the following baseline schemes:
\begin{itemize}
    \item No IRS: UAV serves as a flying relay without IRS elements, only direct BS-UAV-user links are considered.
    \item Random optimization: The UAV trajectory, IRS phases, and SBS beamforming are generated randomly under feasible constraints.
    \item AO approach: An iterative alternating optimization algorithm is used to update trajectory, beamforming, and IRS phases, assuming perfect knowledge of future channels.
\end{itemize}

\subsection{Simulation Results}
\begin{figure}[t!]
\centering
\includegraphics[width=1\linewidth]{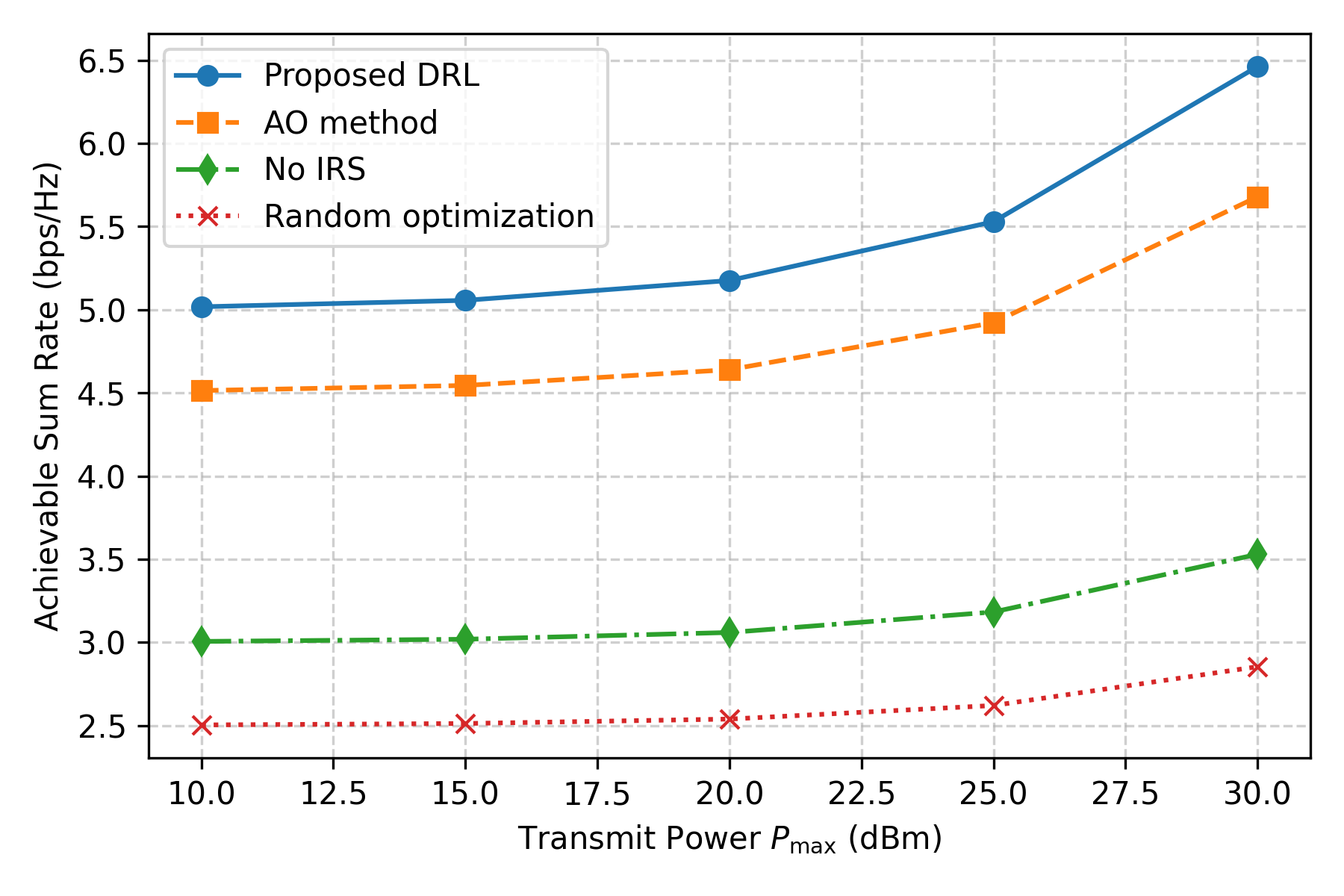}
\vspace{-0.7cm}
\caption{Achievable sum rate versus transmit power.}
\label{fig:rate}
\end{figure}

Fig.~\ref{fig:rate} shows the achievable sum rate of the secondary network as a function of the maximum BS transmit power. The proposed DRL scheme consistently achieves the highest throughput across the entire power range, which confirms its ability to adapt transmission and trajectory decisions to varying link conditions. At $P_{\max}=30$~dBm, the DRL policy provides 25\% gain compared with the AO method, and the advantage becomes more pronounced at lower power levels where intelligent spectrum and beamforming decisions are crucial. In contrast, the no IRS configuration and the random optimization baseline exhibit significant degradation because they fail to provide sufficient spatial degrees of freedom. These results highlight the effectiveness of joint optimization through DRL in fully exploiting both SBS beamforming and UAV-mounted IRS reconfiguration.

\begin{figure}[t!]
\centering
\includegraphics[width=1\linewidth]{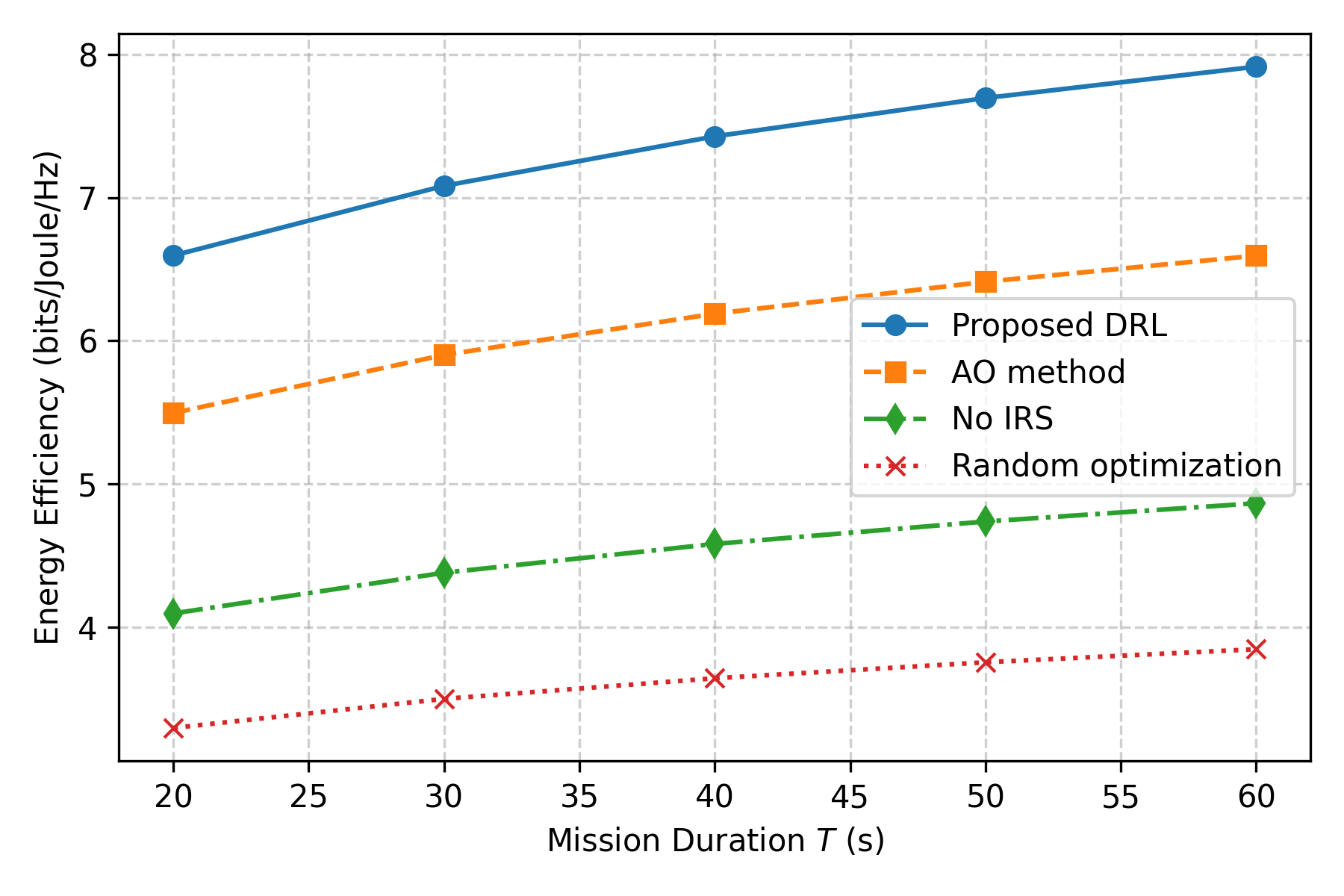}
\vspace{-0.7cm}
\caption{Energy efficiency versus mission duration.}
\vspace{-0.3cm}
\label{fig:ee}
\end{figure}

Fig.~\ref{fig:ee} illustrates the system energy efficiency as a function of UAV mission duration. As the mission time increases, the UAV has more flexibility to adapt its trajectory to favorable channel conditions, which in turn improves EE. The DRL-based approach maintains the highest energy efficiency for all mission lengths, achieving up to 35\% higher EE than AO at long missions and almost doubling the performance compared with random optimization. The improvement stems from the fact that DRL explicitly accounts for the propulsion energy cost when making scheduling and control decisions, enabling a balance between energy consumption and data transmission. The AO method demonstrates reasonable performance but requires heavy computation and cannot adapt to dynamic channel variations in real time. The no IRS baseline is consistently the worst performer, as the absence of reflective gain results in limited spectral efficiency and a poor trade-off between bits delivered and propulsion energy.

\section{Conclusion}\label{sec:conclusion}
This paper has investigated an IRS-assisted UAV secondary network under an OFDM-based spectrum sharing framework, where the UAV-mounted IRS operates as a controllable relay to enhance communication between a multi-antenna BS and ground secondary users while coexisting with a primary network. To maximize the overall energy efficiency, we formulated a joint optimization problem over SBS beamforming, IRS phase shifts, and UAV trajectory, subject to transmit power, passive reflection, kinematic, and interference-protection constraints. The problem is highly non-convex and time-coupled, which renders conventional optimization approaches intractable. To address this challenge, we developed a hybrid DRL-based policy that factorizes discrete subcarrier-user scheduling and continuous radio-and-motion controls, supported by analytic projections and explicit constraint penalties. Simulation results have demonstrated that the proposed framework achieves significant gains in energy efficiency. 

\bibliographystyle{IEEEtran}

\bibliography{IEEEabrv,reference.bib}

\end{document}